\documentclass[aip,jcp,reprint]{revtex4-1}
\usepackage{amsmath}
\usepackage{amsfonts}
\usepackage{amssymb}
\usepackage{mathrsfs}
\usepackage{txfonts}
\usepackage{braket}
\usepackage[version=3]{mhchem}
\usepackage{graphicx}
\usepackage{color}
\usepackage{upgreek}
\usepackage{bm}
\usepackage{multirow}
\usepackage{natbib}
\usepackage{here}
\graphicspath{{./figures/}}

\begin{document}

\title{
  Fluctuation-induced acceleration of inter-ligand exciton transfer in bis(dipyrrinato)Zn(II) complex
}

\author{Hiroki Uratani}
\email{uratani@moleng.kyoto-u.ac.jp}
\affiliation{Department of Molecular Engineering, Graduate School of Engineering, Kyoto University, Kyoto, 615-0510, Japan}
\affiliation{PRESTO, Japan Science and Technology Agency, Kawaguchi, Saitama, 332-0012, Japan}

\author{Hirofumi Sato}
\affiliation{Department of Molecular Engineering, Graduate School of Engineering, Kyoto University, Kyoto, 615-0510, Japan}
\affiliation{Fukui Institute for Fundamental Chemistry, Kyoto University, Kyoto, 606-8103, Japan}


\begin{abstract}
Exciton transfer dynamics between chromophores depends on excitonic coupling,
which is governed by relative orientation between the chromophores.
While the excitonic coupling is treated as a static parameter in many cases,
structural dynamics can introduce time-dependence on the excitonic coupling.
However, influence of the dynamics of excitonic coupling on the exciton transfer has been
scarcely understood.
In the present study, exciton transfer under dynamical fluctuation in excitonic coupling
was investigated via combined use of non-adiabatic molecular dynamics simulations,
exciton density analysis, and a simple two-state model,
for inter-ligand exciton transfer in bis(dipyrrinato)Zn(II) as the example case.
The reaction coordinate for the exciton transfer was obtained {\it a posteriori} via regression analysis
where the target and explanatory variables are diabatic energy gaps and atomic displacements, respectively.
The results suggest that dynamical angular fluctuation between the two dipyrrinato ligands breaks the symmetry
to incidentally increase the excitonic coupling, accelerating the exciton transfer between the ligands.
\end{abstract}

\maketitle

\section{INTRODUCTION} \label{sec_introduction}
Exciton transfer between chromophores plays crucial roles in many areas of photochemistry,
including biological light harvesting,\cite{Krueger1998,Damjanovi2002,Yang2024}
design of light-driven molecular motors,\cite{Pfeifer2020,Liu2025}
and hyperfluorescence strategy to satisfy both efficiency and
color tunability in organic light-emitting diodes.\cite{Deori2024}
According to Fermi golden rule (FGR), in the weak-coupling regime,
the rate of exciton transfer is proportional to $\left|V\right|^2$,
where $V$ is excitonic coupling defined as
\begin{align}
V = \Braket{\Psi_{\rm i}|\hat{H}|\Psi_{\rm f}} \label{coupling}
\end{align}
where $\Psi_{\rm i}$ and $\Psi_{\rm f}$ are the initial and final states of the exciton transfer,
and $\hat{H}$ is Hamiltonian of the system.\cite{Scholes1994}
It is well understood, from both theoretical and experimental perspectives,
that $V$ strongly depends on the relative orientation between the chromophores
that involve the exciton transfer.\cite{Kasha1965,Osuka1988,Saigusa1995,Krueger1998,Wong2004,Zhong2019}
Hence, efforts have been made to control or accelerate the exciton transfer
by tuning $V$ on the basis of molecular- or aggregate-structure design.\cite{Yassar1995,Haldar2019,Imahori2021,Sebastian2021}
From this point of view, $V$ is seen as a static parameter that is determined by the chemical structure of material.
In contrast, because the orientation of chromophores is subject to structural dynamics at finite temperature,
$V$ can vary with time reflecting the change in structure.\cite{Arago2015}
In fact, dynamical fluctuation in relative orientation of chromophores and the resulting distribution of
electronic coupling have been reported in covalent organic frameworks.\cite{KitohNishioka2017,Leo2024}
Taking the dynamical effects in $V$ into account, the exciton transfer dynamics can, in principle,
differ from that is expected from $V$ at the static geometry.
However, the understanding of the effects of dynamic $V$ on exciton transfer phenomena is still very limited
from both points of view of qualitative mechanism interpretation and quantitative estimation of its impact.

\begin{figure}[h]
\centering
\includegraphics[width=3in, bb=0 0 238 110]{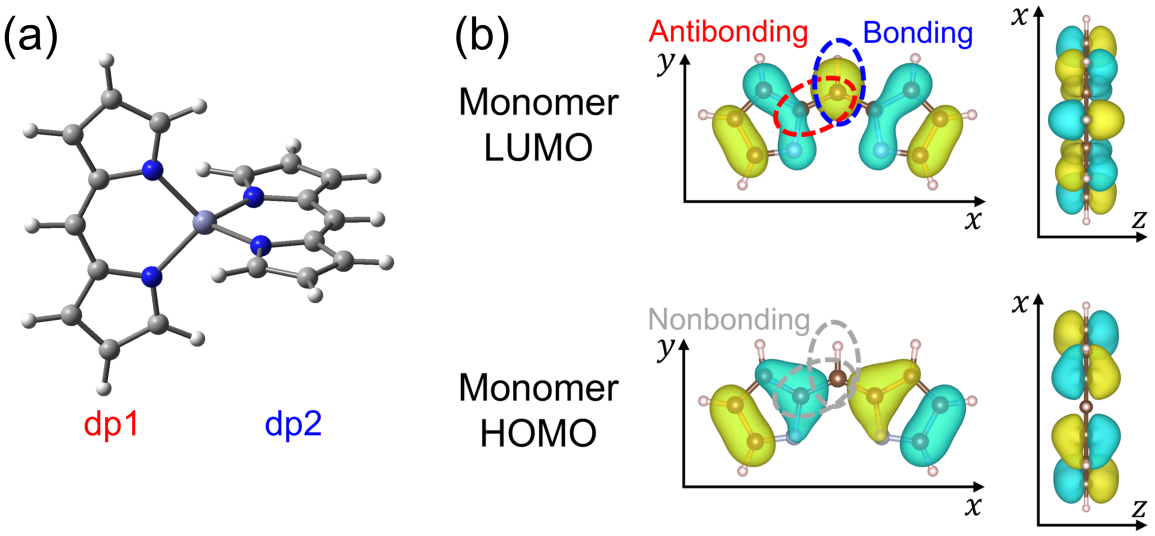}
\caption{
(a) Struture of \ce{Zn(dp)2}. The white, gray, blue, and silver spheres represent H, C, N, and Zn atoms, respectively.
(b) HOMO (lower) and LUMO (upper) of dp monomer anion.
The orbitals have nonbonding, bonding, and antibonding characters with respect to
the bonds highlighted by gray, blue, and red dashed circles, respectively.
Visualized using VESTA software\cite{Momma2011} (isolevel=0.03).}
\label{FIG1}
\end{figure}

In this context, the exciton transport in multinuclear dipyrrin complexes
synthesized by Toyoda, Sakamoto and co-workers\cite{Toyoda2025,Mustafar2025,Matsuoka2015,Tsuchiya2014}
provide much implication for us.
These complexes constitute repetition of the unit \ce{Zn(dp)2} (Figure \ref{FIG1}a), where dp denotes dipyrrin chromophore.
At the ground-state stable geometry, the two dps coordinated to Zn are orthogonal to each other,
where the excitonic coupling between the dps is exactly zero due to the symmetry;\cite{Telfer2011}
in principle, the exciton transfer between these two dps is unable to occur.
On the contrary, it has been experimentally suggested that 
the exciton transfer between these two dps constitutes one of the exciton transfer paths.\cite{Toyoda2025,Tsuchiya2014}
In Ref.~\citenum{Tsuchiya2014} the lower limit of rate constant for the exciton transfer is
estimated as $5\times10^{10}\,{\rm s^{-1}}$ based on the fluorescence quantum yield and lifetime of the dp chromophores.
These results make us expect that the dynamical angular fluctuation between the two dps
breaks the symmetry to yield the nonzero excitonic coupling, enabling the ultrafast exciton transfer.

In the present study, the mechanism of exciton-transfer acceleration via the enhanced $V$ under
the symmetry-breaking dynamical fluctuation
is clarified by the combined use of the non-adiabatic molecular dynamics (NA-MD) simulations,
the exciton density analysis,
and a simplified picture of exciton transfer based on a two-state model.
The results demonstrate that $V$, which is zero at the ground-state stable geometry,
becomes to have finite values via the dynamical symmetry-breaking fluctuation in the dihedral angle between the two dps,
realizing ultrafast exciton transfer.
The atomic displacement mode that works as the reaction coordinate is determined via regression analysis,
clarifying that the dynamics along the reaction coordinate, which drives the exciton transfer,
has the faster timescale compared to the fluctuation in the dihedral angle between the two dps.
These results suggest the dynamical picture of exciton transfer phenomena in the present system, that is,
the slow fluctuation in the dihedral angle between the two dps ``turns on/off'' the exciton transfer,
which is driven by the fast nuclear motion along the reaction coordinate.

\section{METHODS}
\subsection{Non-adiabatic molecular dynamics simulations}
The NA-MD simulations were performed for \ce{Zn(dp)2} complex (Figure \ref{FIG1}a).
Hereafter, we use the labels dp1 and dp2 to distinguish the two dp ligands in \ce{Zn(dp)2}.
Among several NA-MD strategies distinct in the levels of approximation to the quantum nature of nuclei,\cite{CrespoOtero2018}
surface hopping (SH)\cite{Tully1971,Subotnik2016} approaches,
in which the nuclear dynamics are treated as classical trajectories
that can hop among different potential energy surfaces (PESs) to incorporate the non-adiabatic effects,
have been well-established theoretical frameworks to study
exciton transfer dynamics in molecular systems.\cite{Wohlgemuth2020,Gil2021,Stephens2021,Sindhu2022,Mirn2023,Toldo2023,Titov2023,Giannini2024}
In the present study,
the simulations were conducted in SH approach based on the fewest-switches algorithm (FSSH)\cite{Tully1990}
for 100 trajectories in total, where each trajectory was propagated for $250\,{\rm fs}$.
The energy-based decoherence correction\cite{Granucci2007} was applied
with the decay parameter set to the commonly used value (0.1 au). 
The propagation of time-dependent electronic wavefunction and the calculations of non-adiabatic transition probabilities
were performed according to the local diabatization approach\cite{Plasser2012}
with the algorithm modified by Granucci and co-workers in 2019.\cite{AguileraPorta2019}
The nuclear momenta were kept unchanged after each ``frustrated'' hop.
The initial geometries and momenta were randomly sampled from the vibrational Wigner distribution at the electronic ground state at 298.15 K.
The initial electronic state were set to S$_1$ for 50 trajectories and to S$_2$ for other 50 trajectories.
Five electronic excited states (S$_\text{1--5}$) were taken into account for the NA-MD simulations.
The quantum chemical calculations were conducted based on
the linear-response time-dependent density-functional theory (TD-DFT)
using the CAM-B3LYP exchange--correlation functional,\cite{Yanai2004}
which is reported to show generally a good performance in describing excitation energies.\cite{Peach2008}
This is because the CAM-B3LYP is a range-separated hybrid functional\cite{Leininger1997} that is designed to incorporate
the proper amount of long-range exchange potential, which plays a critical role
in the accurate description of excited states.\cite{Iikura2001,Tawada2004}
The Def2SV(P)\cite{Weigend2005} basis set was employed, and the calculations were conducted using
Gaussian 16 program package.\cite{g16}
The NA-MD simulations were performed using Newton-X program\cite{Barbatti2022} interfaced with Gaussian 16.

\subsection{Exciton density analysis}
To characterize the exciton transfer via NA-MD simulations
it is necessary to define the population of exciton on each ligand.
According to the approach developed by Mennucci and co-workers,\cite{Nottoli2018,Yabu2025} at each snapshot,
the exciton density matrix was defined as
\begin{align}
{\bf D}^{\rm ex}_I = \frac{1}{2} \left( {\bf D}^{\rm h}_I + {\bf D}^{\rm p}_I \right)
\end{align}
where $I$ denotes the index of running state at that snapshot.
${\bf D}^{\rm h}_I$ and ${\bf D}^{\rm p}_I$ denotes the attachment and detachment density matrices,\cite{HeadGordon1995}
respectively, corresponding to the excitation from S$_1$ to the state $I$.
Using the exciton density matrix, the exciton Mulliken population on atom $A$ can be defined as
\begin{align}
q^{\rm ex}_A = \sum_{\mu \in A} \left( {\bf D}^{\rm ex}_I {\bf S} \right) \label{ex_mulliken}
\end{align}
The exciton population on dp1, $Q^{\rm ex}_{\rm dp1}$, is defined as the sum of $q^{\rm em}_A$
over the atoms belonging to dp1
\begin{align}
Q^{\rm ex}_{\rm dp1} = \sum_{A \in {\rm dp1}} q^{\rm ex}_A \label{ex_population}
\end{align}
and $Q^{\rm ex}_{\rm dp2}$ is defined in the same way.
Because the number of exciton in the system is unity and the exciton population on \ce{Zn^2+} is negligibly small,
the following relationship is satisfied for the entire simulation time.
\begin{align}
Q^{\rm ex}_{\rm dp1} + Q^{\rm ex}_{\rm dp2} \approx 1
\end{align}
In particular, the local excited (LE) state on dp1 is characterized by
\begin{align}
Q^{\rm ex}_{\rm dp1} \approx 1, \quad Q^{\rm ex}_{\rm dp2} \approx 0
\end{align}
and the LE state on dp2 is characterized by the similar way.
In addition to the LE, the excited states can have the charge-transfer (CT) character from dp1 to dp2 or vice versa.
To quantify the CT character of the excited state, the CT index, $Q^{\rm CT}$, is defined as
\begin{align}
Q^{\rm CT} = \frac{1}{2} \max \left(
\left| Q^{\rm h}_{\rm dp1} - Q^{\rm p}_{\rm dp2} \right|,
\left| Q^{\rm h}_{\rm dp2} - Q^{\rm p}_{\rm dp1} \right|
\right)
\end{align}
where $Q^{\rm h/p}_{\rm dp1/dp2}$ are the Mulliken hole/particle populations on dp1/dp2
defined in the analogy of eqs \ref{ex_mulliken} and \ref{ex_population} but using
$D^{\rm h/p}_I$ instead of $D^{\rm ex}_I$.
The $Q^{\rm CT}$ quantifies the number of electrons transferred between dp1 and dp2;
when the electronic state has the pure CT character, i.e., exactly one electron is
transferred from dp1 to dp2 (or from dp2 to dp1), then $Q^{\rm CT}=1$.
As the opposite limit, if the state is completely characterized as LE, then $Q^{\rm CT}=0$.
In general, the excited states can have both LE and CT characters partially,
and to classify the states one has to identify the character of majority.
In the present analyses, the electronic state was classified
as the LE on dp1 when $Q^{\rm ex}_{\rm dp1}>Q^{\rm ex}_{\rm dp2}$ and $Q^{\rm CT}<0.5$,
as the LE on dp2 when $Q^{\rm ex}_{\rm dp1}<Q^{\rm ex}_{\rm dp2}$ and $Q^{\rm CT}<0.5$,
and as the CT when $Q^{\rm CT}>0.5$.
An exciton transfer event is defined as a change in the magnitude relationship
between $Q^{\rm ex}_{\rm dp1}$ and $Q^{\rm ex}_{\rm dp2}$ under the condition that the system is at the LE states.
The exciton density analysis presented here conceptually resembles to the charge transfer number
introduced by Plasser and Lischka\cite{Plasser2012},
which is defined on the basis of transition density matrix.
Instead of the transition density matrix,
the present analysis is grounded on the relaxed difference density matrix,
which fully accounts for the orbital relaxation upon the electronic excitation.\cite{Herbert2024}

\subsection{Two-state model for exciton transfer}
The NA-MD results were interpreted according to the following two-state model\cite{Nitzan2006Book,Landry2012}
that describes the exciton transfer.
The diabatic energies for LE states of dp1 ($E_{\rm dp1}$) and dp2 ($E_{\rm dp2}$) are described in quadratic forms
\begin{align}
E_{\rm dp1/dp2}\left(Q\right) &= \frac{1}{2}k \left(Q \pm Q_0\right)^2 \label{quadratic}
\end{align}
where $Q$ denotes the atomic displacement in the direction of reaction coordinate ${\bf Q}$, i.e., $Q = \left|{\bf Q}\right|$.
Because dp1 and dp2 are chemically equivalent, they share the same value of force constant $k$.
The diabatic energies are crossed with each other at $Q=0$, and the diabatic energy minima are located at
$Q=-Q_0$ (for dp1) and $Q=Q_0$ (for dp2).
From eq.~\ref{quadratic}, the diabatic energy gap $\Delta E$ linearly depends on $Q$
\begin{align}
\Delta E\left(Q\right) = E_{\rm dp1}\left(Q\right) - E_{\rm dp2}\left(Q\right) = 2kQ_0Q \label{linear_eq}
\end{align}
Within the Hilbert space spanned by these two states, the Hamiltonian is written as
\begin{align}
{\bf H}\left(Q\right) =
\begin{pmatrix}
E_{\rm dp1}\left(Q\right) & V\left(\phi\right) \\
V\left(\phi\right) & E_{\rm dp2}\left(Q\right)
\end{pmatrix}
\end{align}
where $V\left(\phi\right)$ denotes the excitonic coupling, which is assumed as a function of
the dihedral angle $\phi$ between the two dps; this assumption is to be validated in \ref{phi_and_V}.
By diagonalizing the Hamiltonian, the adiabatic energy levels are obtained as
\begin{align}
E_{\rm S_1/S_2}\left(Q\right) = \frac{1}{2}kQ^2 + \frac{1}{2}kQ_0^2 \mp \sqrt{k^2Q_0^2Q^2+V\left(\phi\right)^2}
\end{align}
At the diabatic crossing point, i.e., $Q=0$, the adiabatic energy gap is simply written as
\begin{align}
E_{\rm S_2}\left(0\right) - E_{\rm S_1}\left(0\right) = 2\left|V\left(\phi\right)\right| \label{2v}
\end{align}

\section{RESULTS AND DISCUSSION}
\subsection{Characterization of low-lying excited states}
Figure \ref{FIG1}b shows the HOMO and LUMO of an isolated, anionic dp ligand
extracted from the S$_0$-optimized geometry of \ce{Zn(dp)2}.
In the $x$-axis, the HOMO and LUMO are odd and even functions, respectively.
On the other hand, reflecting the ${\rm \pi}$-character of these orbitals,
both the HOMO and LUMO are odd functions in the $z$-axis.
The lowest LE wavefunction of a dp monomer,
which is characterized as the HOMO$\rightarrow$LUMO excitation,
has the odd (odd$\times$even) and even (odd$\times$odd) symmetry in $x$ and $z$ directions, respectively.
When the relative orientation of two dps in $\ce{Zn(dp)2}$ is strictly orthogonal,
$\Psi_{\rm i}$ and $\Psi_{\rm f}$ in eq.~\ref{coupling} are odd and even, respectively, or even and odd, respectively,
for $x$ and $z$ directions.
Because $\hat{H}$ is total symmetric, the right-hand side of eq.~\ref{coupling} is zero,
annihilating the excitonic coupling between the two dps in \ce{Zn(dp)2} at the ground-state stable geometry.

The calculated excitation energies and oscillator strengths of \ce{Zn(dp)2} are listed in Table \ref{low-lying}.
The calculations were conducted on the ground-state optimized geometry,
where the dihedral angle between dp1 and dp2 is 90$^\circ$.
The strong oscillator strengths of S$_1$ and S$_2$ states suggest that these states are
composed of the linear combinations of local excited (LE) states on dp1 and dp2.
The degeneracy of these states reflects the fact that $V$ equals to zero when the two dps are orthogonal to each other.
In addition, the degenerated S$_3$ and S$_4$ states, which are 0.458 eV above S$_1$ and S$_2$ states,
have far smaller oscillator strengths, suggesting the charge-transfer (CT) character.
The above electronic-state characterization is consistent with the natural transition orbitals\cite{Martin2003}
presented in Figure S1 in Supporting Information.
These results suggest that the exciton transfer between dps is described within the Hilbert space spanned by
the low-lying two bright adiabatic states.
\begin{table}[h]
\centering
\caption{Excitation energies and oscillator strengths of \ce{Zn(dp)2} at the S$_0$-optimized geometry.}
\label{low-lying}
\begin{tabular}{ccc} \hline \hline
State & Excitation energy [eV] & Oscillator strength \\ \hline
S$_4$ & 3.8077 & 0.0000 \\
S$_3$ & 3.8077 & 0.0000 \\
S$_2$ & 3.3506 & 0.6150 \\
S$_1$ & 3.3506 & 0.6150 \\ \hline \hline
\end{tabular}
\end{table}

\subsection{Temporal changes in electronic-state populations}
\begin{figure}
\centering
\includegraphics[width=3.3in, bb=0 0 238 346]{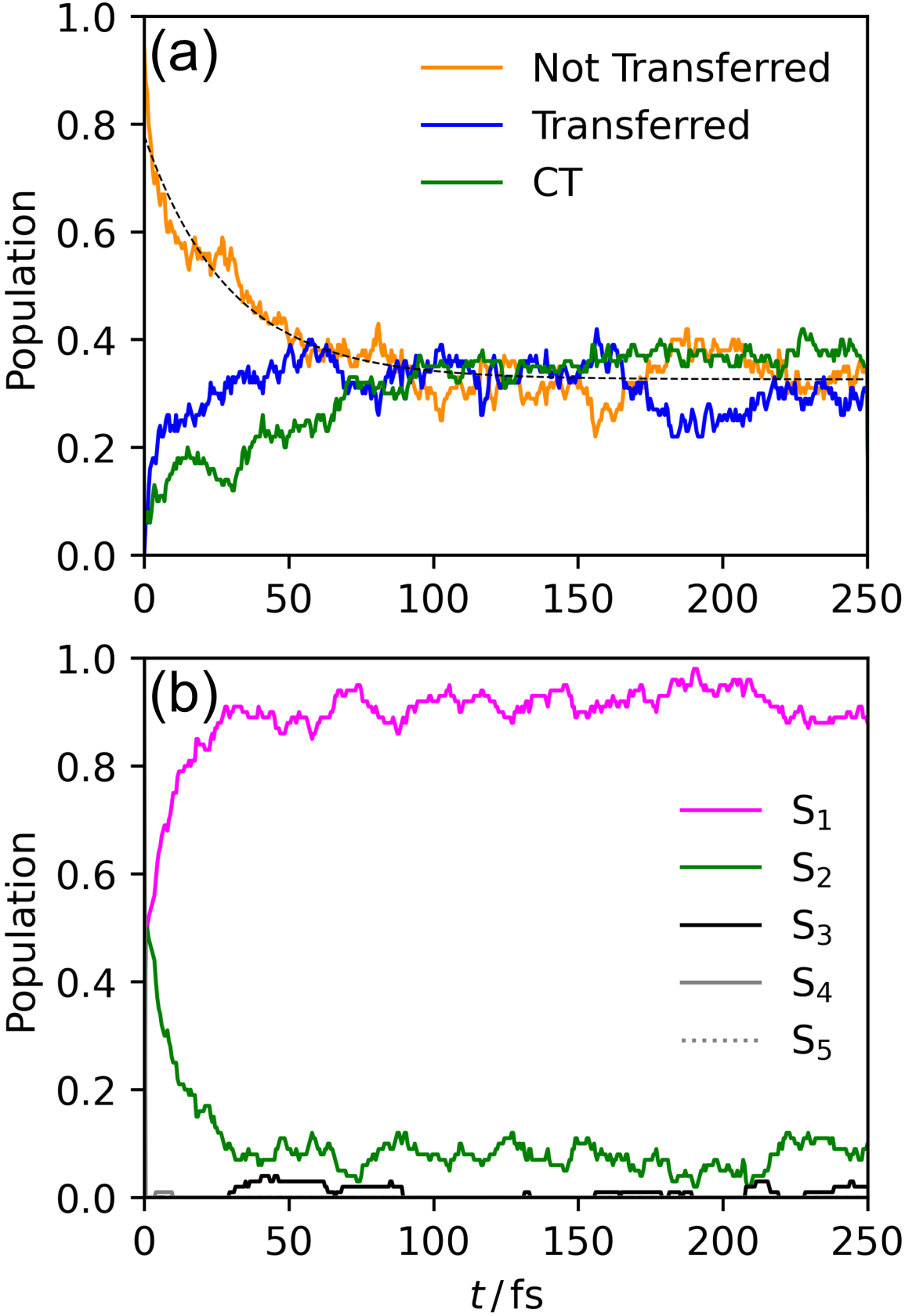}
\caption{
(a) Temporal changes in electronic state populations.
Orange, blue, and green lines indicate the populations of electronic states
classified as LE (Not Transferred), LE (Transferred), and CT.
Dashed black line indicates the fitting curve according to eq.~\ref{fitting}.
(b) Temporal changes in adiabatic state populations.
}
\label{FIG2}
\end{figure}

Figure \ref{FIG2}a presents the temporal changes in electronic-state populations
obtained from the NA-MD simulations averaged over 100 trajectories.
Here, the electronic states are labelled as LE (Not Transferred), LE (Transferred), or CT,
and the population of each state is defined as the number fraction of trajectories assigned to that state.
At the beginning of each trajectory, i.e., $t=0$, the system was at LE on either dp1 or dp2.
At $t>0$, when the system is at LE on dp that the exciton is located at $t=0$,
the electronic state is classified as LE (Not Transferred).
In contrast, when the system is at LE on the other dp, the state is classified as LE (Transferred).
The decay of LE (Not Transferred) and the rise of LE (Transferred) in Figure \ref{FIG2}a
are in the time scale of $10^2\,{\rm fs}$, indicating the ultrafast intramolecular exciton transfer.
As expected by the fact that the two dps in the system are chemically equivalent,
LE (Not Transferred) and LE (Transffered) converges to the same value,
i.e., the probabilities of finding an exciton on dp1 and dp2 become identical in the long time limit.
In addition, the population of CT converges to the similar value, being consistent with the fact that
the geometry-relaxed CT state is only slightly (25 meV) higher in potential energy than the geometry-relaxed LE state
(Table S1).
The time constant $\tau$ for the exciton transfer was estimated by fitting
the time series of ``Not Transferred'' population $p\left(t\right)$ with
\begin{align}
p\left(t\right) = A\exp\left(-t/\tau\right) + B \label{fitting}
\end{align}
where $A$, $B$, and $\tau$ are the fitting parameters.
The result is $\tau=3\times10^{-14}\,{\rm s}$, successfully falling onto a shorter value
than the experimentally suggested upper limit of the time constant
($1/(5\times10^{10}\,{\rm s^{-1}})=2\times10^{-11}\,{\rm s}$).\cite{Tsuchiya2014}
Figure \ref{FIG2}a also shows that the formation of CT states occurs in a time scale
that is close to but slower than that of intramolecular exciton transfer.

The temporal changes of state populations in the adiabatic representation are plotted in Figure \ref{FIG2}b,
suggesting that the majority of trajectories run S$_1$,
and the population of the states higher than S$_2$ is negligible.
The population difference between S$_1$ and S$_2$ can be understood from the viewpoint of potential energy difference;
while S$_1$ and S$_2$ are degenerate at the S$_0$-optimized geometry (Table \ref{low-lying}),
the structural fluctuation can break the symmetry to lift the degeneracy of these two states.
The averaged gap between the S$_1$ and S$_2$ obtained from the NA-MD simulations was 259 meV,
being consistent with the significant difference in populations.

\subsection{``Adiabatic'' and ``non-adiabatic'' pathways of exciton transfer}
\begin{figure}
\centering
\includegraphics[width=3.3in, bb=0 0 238 99]{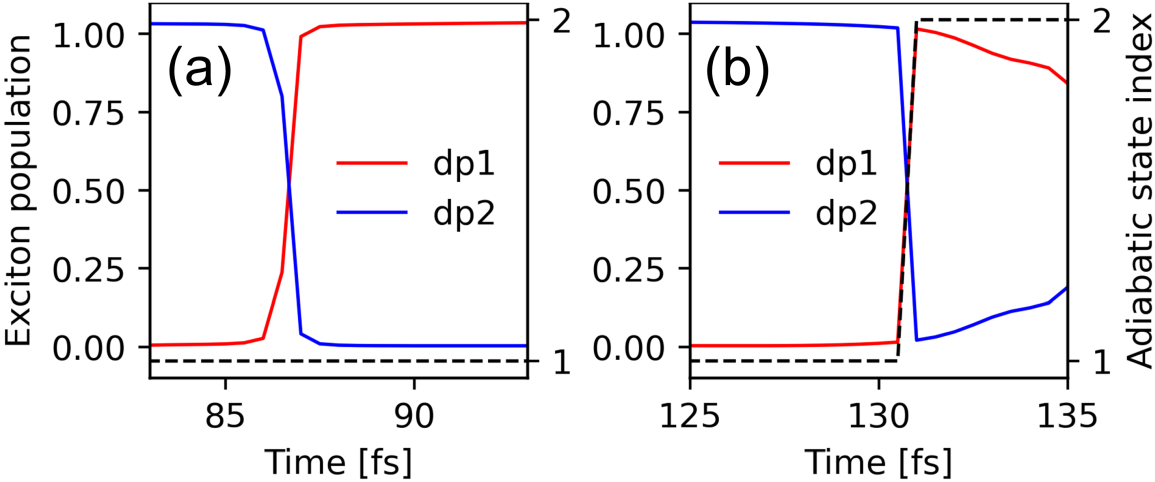}
\caption{
(a) Temporal changes of $Q^{\rm ex}_{\rm dp1}$ and $Q^{\rm ex}_{\rm dp2}$ for two example trajectories
that represent ``adiabatic'' exciton transfer pathway and (b) that represent ``non-adiabatic'' exciton transfer pathway.}
\label{FIG2-1}
\end{figure}

\begin{figure}
\centering
\includegraphics[width=2.0in, bb=0 0 238 225]{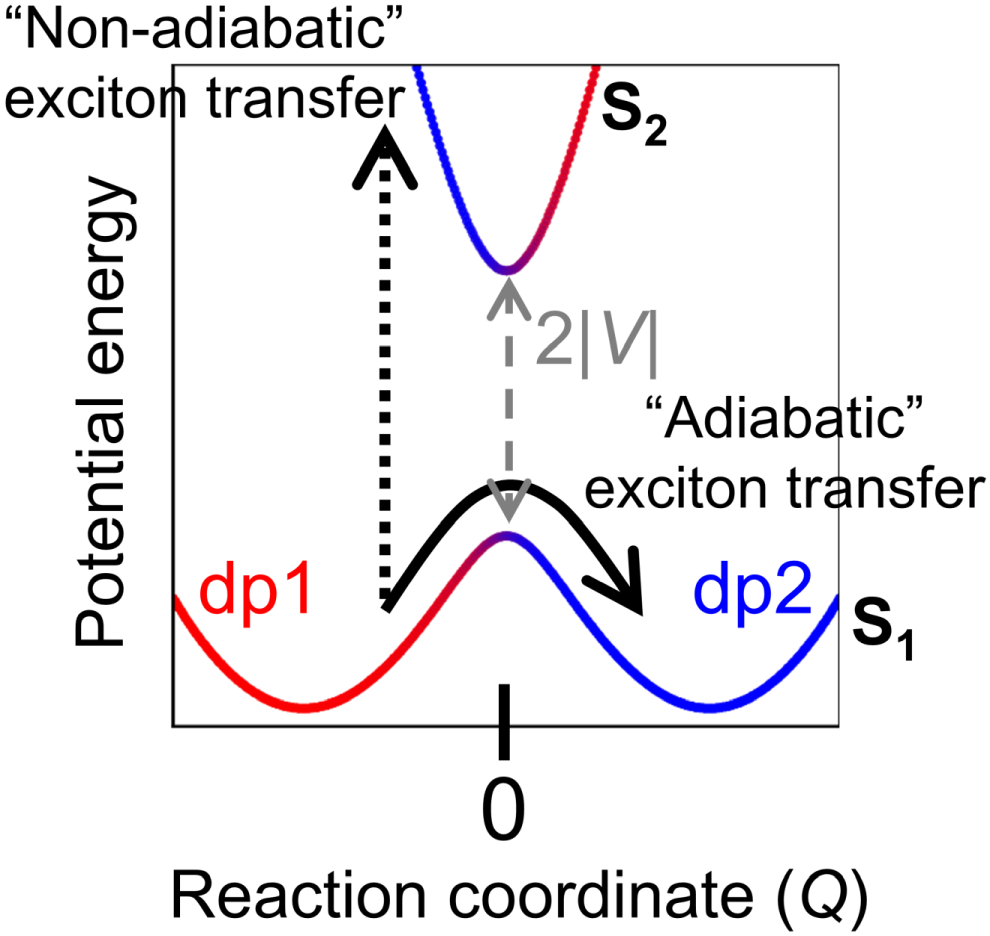}
\caption{Schematic illustration of ``adiabatic'' and ``non-adiabatic'' exciton transfer pathways
on adiabatic potential energy surfaces (PESs).
The red and blue colors of PESs indicate that the exciton is dominantly localized on dp1 and on dp2, respectively.
}
\label{adiabatic_nonadiabatic}
\end{figure}

Figures \ref{FIG2-1}a and \ref{FIG2-1}b illustrate the temporal changes of
exciton populations on each dp, that is, $Q^{\rm ex}_{\rm dp1}$ and $Q^{\rm ex}_{\rm dp2}$.
Note that the total of exciton population is not exactly unity because of
the contribution of de-excitation in the TD-DFT calculations.
The results are shown for two example trajectories which represents distinct characters of exciton transfer processes
that we hereafter call ``adiabatic'' and ``non-adiabatic'' mechanisms.
In ``adiabatic'' mechanism (Figure \ref{FIG2-1}a), the temporal changes in exciton populations of the two dps
are, in principle, continuous.
The physical interpretation of this process is presented in Figure \ref{adiabatic_nonadiabatic}.
The LE states on dp1 and on dp2 are represented by two distinct diabatic surfaces,
which are red- and blue-colored in Figure \ref{adiabatic_nonadiabatic}.
In the adiabatic picture, the potentials are represented by the lower (S$_1$) and the upper (S$_2$)
adiabatic surfaces.
Each of these adiabatic states are constituted by a linear combination of the two diabatic states,
where the dominant diabatic state switches at the geometry of diabatic crossing ($Q=0$),
as depicted by the change in the line color.
When the system pass through $Q=0$ adiabatically, i.e., without a non-adiabatic transition, 
the dominant diabatic state switches, which means that an exciton is transferred
from one of the two dps to the other dp.
Another type of the exciton transfer process is called ``non-adiabatic'' mechanism (Figure \ref{FIG2-1}b),
where the exciton populations of the two dps discontinuously changes in association to a non-adiabatic transition.
This mechanism is also depicted in Figure \ref{adiabatic_nonadiabatic};
when a non-adiabatic transition occur at a position far from $Q=0$,
the transition entails the switching of dominant diabatic state.
Among all the exciton tranfer events over 100 trajectories, the fractions of ``adiabatic'' and ``non-adiabatic''
exciton transfer were 74.9\% and 25.1\%, respectively,
suggesting that ``adiabatic'' mechanism is the dominant exciton transfer pathway.
In addition, it is worth to comment on the difference in recrossing timescales between the ``adiabatic'' and ``non-adiabatic''
exciton transfer pathways.
Figure S2 shows the distribution of the time between a hopping and the next, i.e., backward, hopping.
In the case of ``adiabatic'' exciton transfer, a sharp peak around 2.5 fs
and a broad peak around 20 fs can be found, where the former may correspond to the coherent recrossing.
On the other hand, in the case of ``non-adiabatic'' exciton transfer, the second peak is less clear,
implying that the coherent recrossing is predominant.

\subsection{Enhanced excitonic coupling via symmetry-breaking dihedral angle fluctuation} \label{phi_and_V}
\begin{figure}[h]
\centering
\includegraphics[width=3.3in, bb=0 0 504 502]{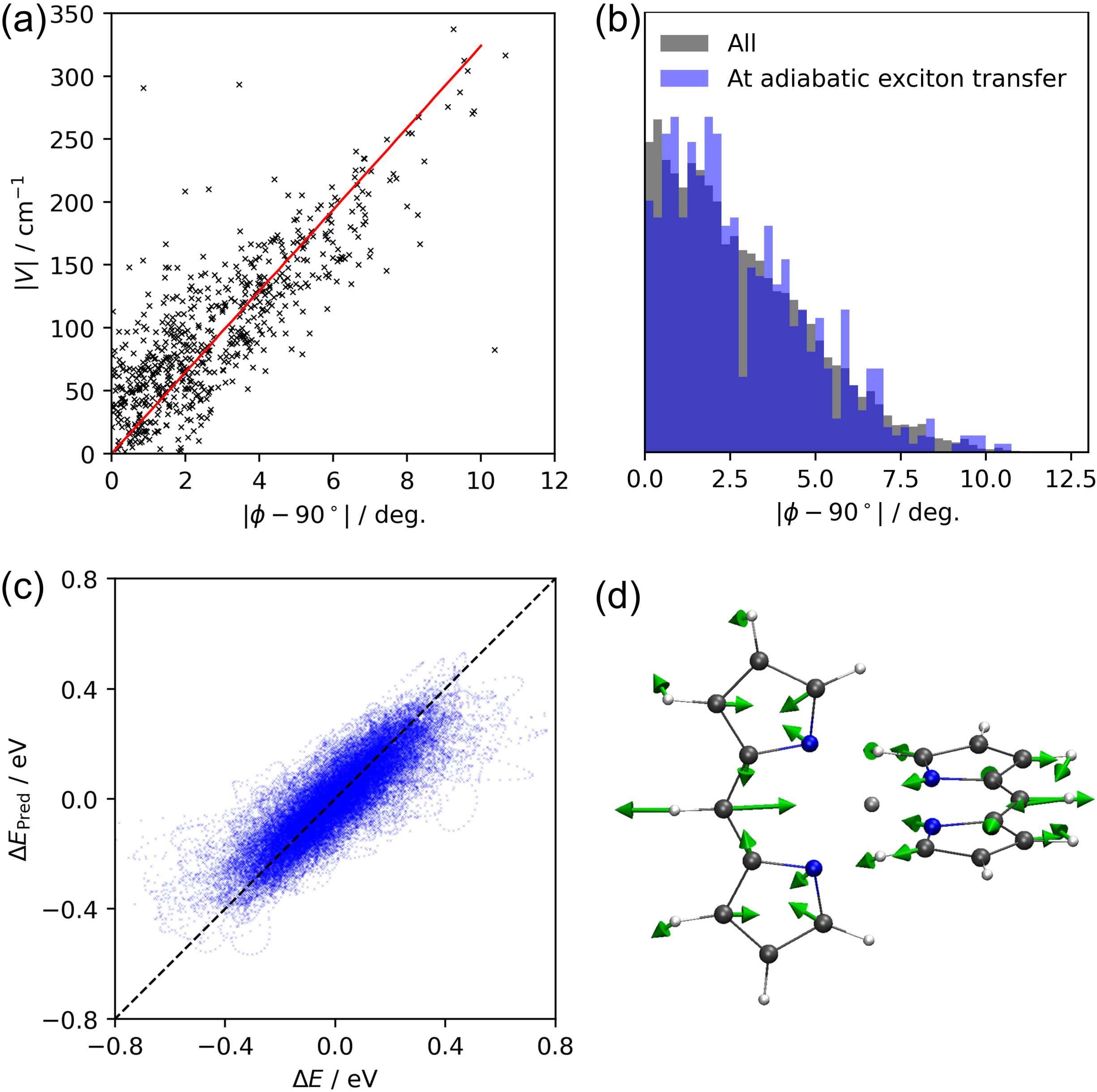}
\caption{
(a) Absolute values of excitonic coupling ($|V|$) versus displacement of the dihedral angle
between the two dps ($|\phi-90^\circ|$) at the moments of ``adiabatic'' exciton transfer events (black).
$|V|$ obtained via single-point TD-DFT calculations with varying $\phi$ is also indicated (red line).
(b) Histograms of the dihedral angle between the two dps ($\phi$) for all snapshots (gray) and
for snapshots at ``adiabatic'' exciton transfer events (blue).
(c) Comparison of the diabatic energy gaps obtained from the ``monomer'' TD-DFT calculations ($\Delta E$),
and those predicted from the linear regression model ($\Delta E_{\rm Pred}$).
(d) Atomic displacement vector in the direction of $\bar{\bf Q}$ (green arrows).
White, gray, blue, and silver spheres represent H, C, N, and Zn atoms, respectively.}
\label{FIG4}
\end{figure}

The absolute values of excitonic coupling ($|V|$) are plotted against the dihedral angle
are plotted against the dihedral angle $\phi$ between the two dps (Figure \ref{FIG4}a, red solid line).
The detailed definition of $\phi$ is presented in Supporting Information S2.
Here, the values of $|V|$ were obtained as the half of adiabatic energy gaps between S$_1$ and S$_2$
estimated via single-point TD-DFT calculations with varying $\phi$ from the S$_0$-optimized geometry.
The result indicates that $|V|$ has a good correlation with $|\phi-90^\circ|$.
In addition, Figure \ref{FIG4}a also presents the values of $|V|$ at the ``adiabatic'' exciton transfer events
with black cross marks, suggesting that the single-point result explains these distribution well.
Note that the ``adiabatic'' exciton transfer can occur at no positions other than the diabatic crossing point;
$|V|$ can be calculated as the half of adiabatic energy gap at an ``adiabatic'' exciton transfer event.

In addition, Figure \ref{FIG4}b indicates that $|\phi-90^\circ|$ has distribution
over the range of approximately $0^\circ$ to $10^\circ$.
While the overall distribution (gray) has the peak around $0^\circ$,
at ``adiabatic'' exciton transfer events (blue) the peak of distribution is shifted to finite value of $|\phi-90^\circ|$,
suggesting that the exciton transfer becomes more probable when the dihedral angle is deviated from 90$^\circ$.
These results indicate that the dynamic fluctuation in the dihedral angle between the two dps facilitates the exciton transfer
by increasing the excitonic coupling.
Combining the linear dependence of $\left|V\right|$ on $\phi$ (Figure \ref{FIG4}a)
and the probability distribution of $\phi$ (Figure \ref{FIG4}b), the averaged value of $\left|V\right|^2$ over the entire
dynamics ($\Braket{\left|V\right|^2}$) can be calculated as
$\Braket{\left|V\right|^2} = 1.29\times10^4\,{\rm cm^{-2}}$,
leading to the root-mean-square $|V|$ of $113\,{\rm cm^{-1}}$.

\subsection{Reaction coordinates for exciton transfer}
According to eq.~\ref{linear_eq}, the reaction coordinate ${\bf Q}$ is the degree of freedom that has
the linear relationship with the diabatic energy gap $\Delta E = E_{\rm dp1} - E_{\rm dp2}$.
To find such degree of freedom, we first calculated the diabatic energies
$E_{\rm dp1}$ and $E_{\rm dp2}$ for each snapshot geometry
obtained from the NA-MD simulations.
Each of $E_{\rm dp1}$ and $E_{\rm dp2}$ was determined as
the first singlet excitation energy obtained with a TD-DFT calculation
for the ``monomer'' geometry taken from a snapshot geometry.
The ``monomer'' geometry was constructed by substituting one of the dp ligands with \ce{NH3} and \ce{Cl-},
where the detailed procedure is presented in Supporting Information S3.
The direction of reaction coordinate $\bar{\bf Q}={\bf Q}/Q$ was determined via the linear regression,
where the target is $\Delta E$ and the descriptors are the atomic displacements
from the stable geometry at S$_0$ state (${\bf R}_0$).
Namely
\begin{align}
\Delta E^{(i)}_{\rm Pred} = a{\bar{\bf q}}\cdot\left({\bf R}^{(i)} - {\bf R}_0\right) + b
\end{align}
where the superscript `$(i)$' denotes the index of snapshot.
${\bf R}^{(i)}$ is the geometry at the snapshot ${i}$,
and $\Delta  E^{(i)}_{\rm Pred}$ is the predicted value for $\Delta E^{(i)}$.
$\bar{\bf q}$ is a unit vector that has the same dimension with ${\bf R}$.
The slope $a$ and the intercept $b$ are scalars.
$\bar{\bf Q}$ was determined via the least-squares method, that is,
\begin{align}
\bar{\bf Q}, A, B = \underset{\bar{\bf q},a,b} {\operatorname{argmin}}\sum_i\left|\Delta E^{(i)}_{\rm Pred}\left(\bar{\bf q},a,b\right) - \Delta E^{(i)}\right|^2
\end{align}
Here, $A$ and $B$ are optimized values of $a$ and $b$, respectively,
where the intercept $B$ fell onto a negligibly small value ($B=0.0032\,{\rm eV}$).
Figure \ref{FIG4}c compares $\Delta E$ and the predicted value $\Delta E_{\rm Pred}$ for each snapshot geometry,
indicating that $\Delta E$ is well described linearly with the found reaction coordinate ${\bf Q}$.
Note that one can also choose the S$_1$ stable geometry, instead of the S$_0$ stable geometry, as the origin.
The results for this case are presented in Section S7 in Supporting Information,
indicating that the numerical outcomes are almost unchanged regardless of the choice of the origin.

Figure \ref{FIG4}d visualizes the displacement vector corresponding to $\bar{\bf Q}$,
indicating that $\bar{\bf Q}$ is composed of in-plane deformation of dps,
where the atomic displacements in two dps are opposite in phase.
The deformation involves the stretching motion of C--H bonds on the central carbon atom, and
the stretching motion of C--C bonds connecting the central carbon atom and 5-membered rings.
In Figure \ref{FIG1}b, the HOMO and LUMO of a dp monomer are visualized
as well as the C--H and C--C bonds that stretch (dashed circles).
The relationship between $\bar{\bf Q}$ and the diabatic energy gap can be qualitatively understood
in terms of the relative phases in the HOMO and LUMO.
As indicated in Figure \ref{FIG1}b, the HOMO has nonbonding character for both of the C--H and C--C bonds,
while the LUMO has bonding and antibonding characters with respect to the C--H and C--C bonds, respectively.
These features suggest that the stretching motion of the C--H and C--C bonds affects the LUMO level
by changing the amount of bonding and antibonding interactions, whereas it has no significant effect on the HOMO level.
Overall, the atomic displacement along $\bar{\bf Q}$ modulates the HOMO--LUMO gaps of the two dp monomers
in opposite direction, resulting in the change in diabatic energy gap.

\subsection{Characterization of structural dynamics}
\begin{figure}[H]
\centering
\includegraphics[width=3.3in, bb=0 0 238 400]{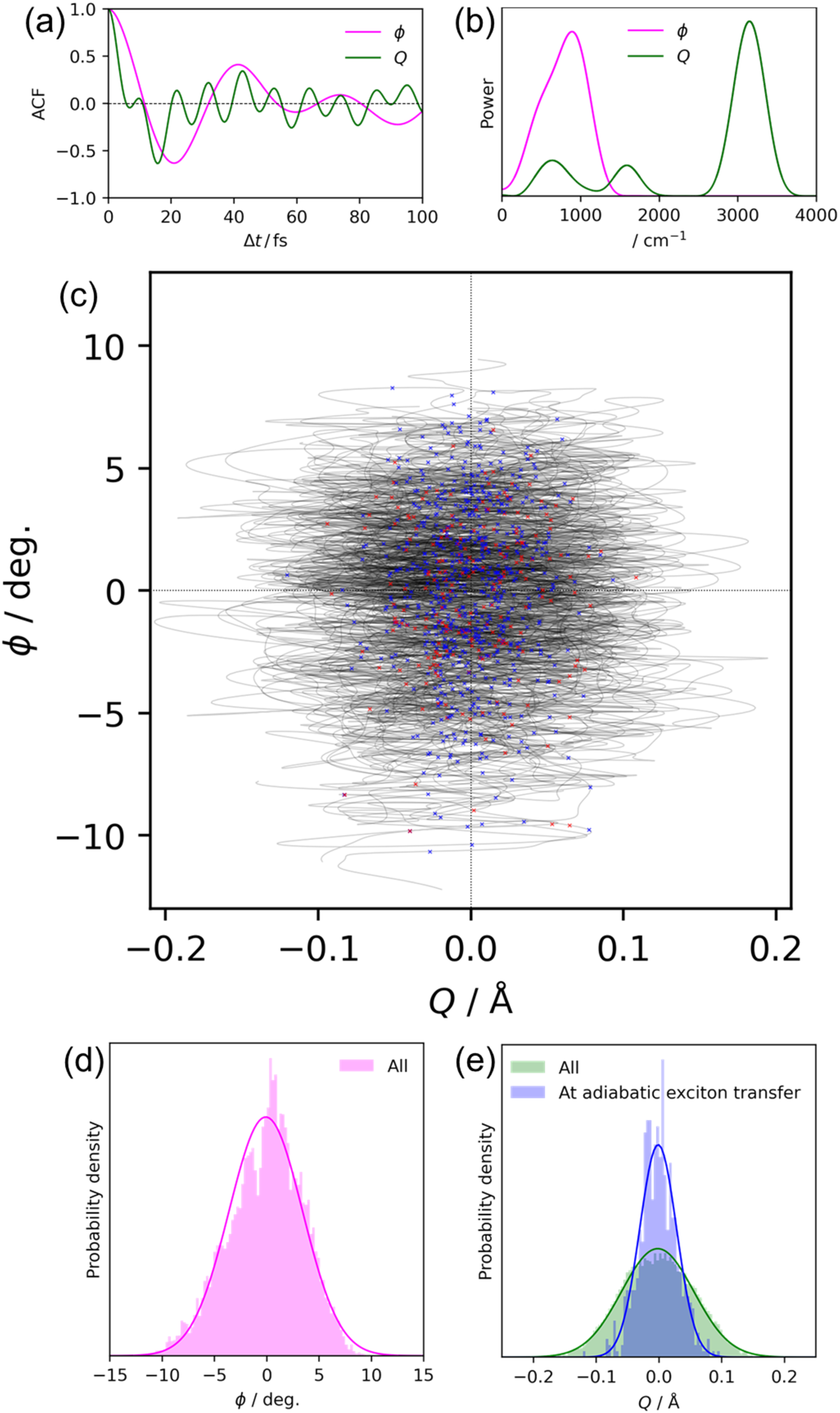}
\caption{
(a) VACFs for the dihedral angle ($\phi$) and the reaction coordinate ($Q$).
(b) Power spectra of VACFs.
(c) 
  Temporal changes of $\phi$ and $Q$ for all NA-MD trajectories.
  The points where ``adiabatic'' and ''non-adiabatic'' exciton transfer events occurred are indicated by
  blue and red cross marks, respectively.
(d) Distribution of $\phi$ sampled from the NA-MD trajectories (green).
(e) Distribution of $Q$ sampled from the NA-MD trajectories and at ``adiabatic'' exciton transfer (blue).
}
\label{FIG5}
\end{figure}

Figure \ref{FIG5}a presents the velocity autocorrelation functions (VACFs) for the dihedral angle $\phi$ and 
the reaction coordinate $Q$ obtained from the NA-MD simulations,
indicating the slow and fast oscillations in the VACFs for $\phi$ and $Q$, respectively.
Figure \ref{FIG5}b shows the power spectra obtained from the Fourier transformation of the squared VACFs,
highlighting the relative timescales of the dynamics in these two degrees of freedom.
The characteristic frequency for $\phi$ dynamics is relatively low (up to $1000\,{\rm cm^{-1}}$),
reflecting the fact that the potential for the dihedral rotation is governed by the soft Zn--N bonds
and nonbonding interactions between two dp planes.
On the contrary, Figure \ref{FIG5}b indicate that the spectrum of $Q$ has three peaks around 800, 1600, and $3100\,{\rm cm^{-1}}$.
The normal-mode decomposition of $\bar{\bf Q}$ (Section S6 in Supporting Information) suggests that these peaks
correspond to the bending motions in the carbon framework,
the stretching motions of C--C and C--N bonds, and the stretching motions of C--H bonds, respectively.
Among the three peaks, the peak around $3100\,{\rm cm^{-1}}$ has the dominant contribution.
These results suggest that the timescale of individual exciton transfer events, which is driven by the dynamics in $Q$, 
is faster than the timescale of fluctuation in $\phi$, which facilitates the exciton transfer by intensifying the
excitonic coupling $V$.
In Figure \ref{FIG5}c, all NA-MD trajectories are mapped onto the ($Q$, $\phi$) plane,
showing the ``horizontal hatching'' pattern, reflecting the relatively fast and slow dynamics of $Q$ and $\phi$, respectively.
Figure \ref{FIG5}c also shows that the dynamics of $Q$ and $\phi$ are independent of each other;
no substantial correlation between $Q$ and $\phi$ is observed.
Figures \ref{FIG5}d and \ref{FIG5}e show the distribution of $\phi$ and $Q$ in the NA-MD trajectories, respectively,
indicating that the distribution is approximately Gaussian.
This result suggests that the dynamics of both $\phi$ and $Q$ are dominated by thermal fluctuation.
Figure \ref{FIG5}e also indicates that $Q$ at the ``adiabatic'' exciton transfer events shows
the substantially narrower distribution than $Q$ in the entire NA-MD trajectories,
suggesting that ``adiabatic'' exciton transfer events tend to occur in the vicinity of $Q=0$.
Note that the corresponding VACFs and the distribution for the ground-state Born--Oppenheimer trajectories
are shown in Figures S7 and S8, respectively, exhibiting the similar features to the NA-MD results (Figure \ref{FIG5}).

\subsection{Summary of suggested exciton transfer mechanism} \label{subsec_2dpes}
\begin{figure}[h]
\centering
\includegraphics[width=3.3in, bb=0 0 238 171]{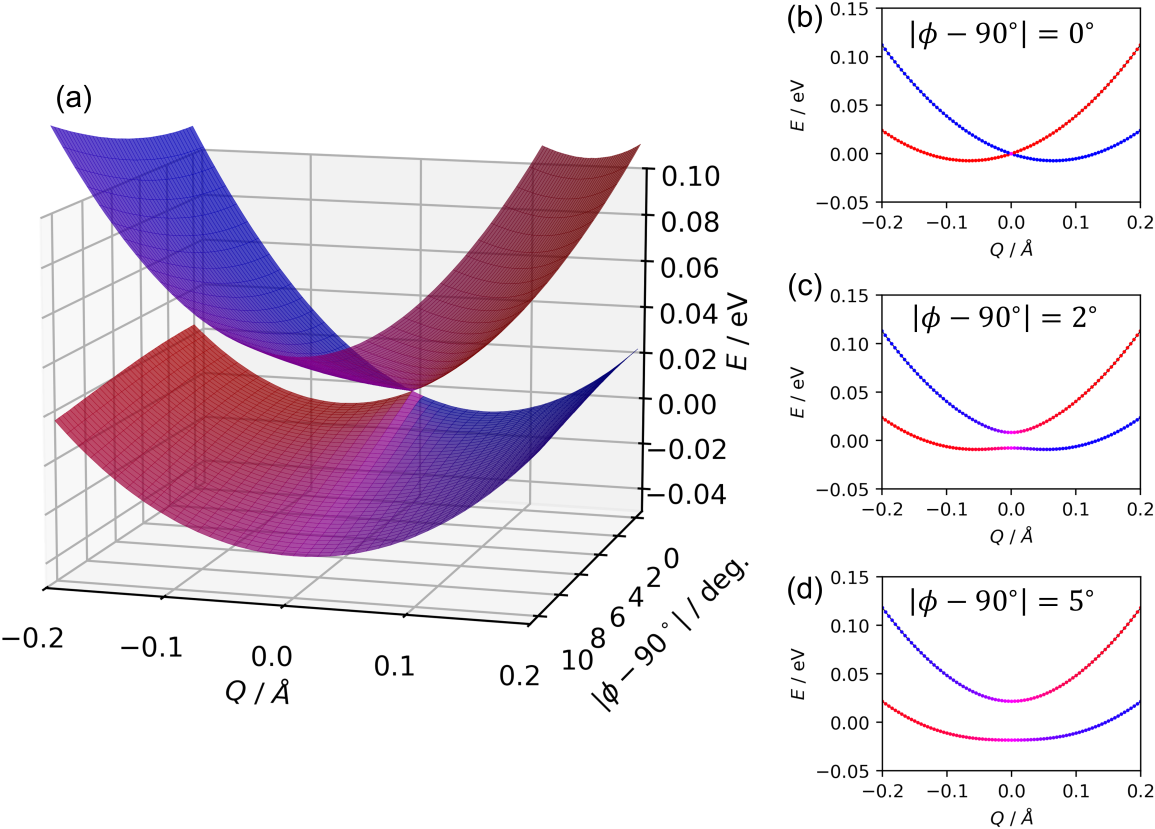}
\caption{
(a) Adiabatic two-dimensional PESs of S$_1$ and S$_2$ states plotted as functions of $Q$ and $\phi$.
Intensities of red and blue colors indicate the exciton populations on dp1 and dp2, respectively.
(b) Corresponding one-dimensional PESs as functions of $Q$, at $\left|\phi-90^\circ\right|$ fixed to
$0^\circ$, (c) to $2^\circ$, and (d) to $5^\circ$.
}
\label{FIG6}
\end{figure}

Figure \ref{FIG6}a presents the adiabatic potential energy surfaces (PESs) of S$_1$ and S$_2$ states
as functions of $Q$ and $\phi$,
obtained from TD-DFT calculations for finite-displaced geometries in these dimensions.
At each point on a PES, the exciton population on dp1 and dp2 were calculated and
these are indicated by the intensity of red and blue colors, respectively.
As discussed in Figures \ref{FIG5}a-c, the temporal change in $Q$ is considerably faster than that of $\phi$,
allowing us to qualitatively describe the exciton transfer on one-dimensional PESs
versus $Q$ with a certain fixed value of $\phi$.
When $\phi$ is close to $90^\circ$ (Figure \ref{FIG6}b), the two PESs are closely contacted at the diabatic crossing point,
where a non-adiabatic transition is very likely to occur, minimizing the probability of exciton transfer.
Deviation of $\phi$ from $90^\circ$ (Figure \ref{FIG6}c) widens the adiabatic gap at the diabatic crossing point,
i.e., increases the excitonic coupling $|V|$,
making the non-adiabatic transition less probable, resulting in higher probability of exciton transfer.
At large deviation of $\phi$ from $90^\circ$ (Figure \ref{FIG6}d), the S$_1$ PES loses two minima and becomes a simple
downward convex function that connects the two diabatic states, i.e., the LE on dp1 and the LE on dp2, 
suggesting that the exciton transfer occurs barrierlessly in this situation.
These results suggest that $\phi$ acts as a switch that turns on/off the exciton transfer,
which is driven by the faster nuclear motion along $Q$.
While the excitonic coupling is zero at the fixed stable geometry ($\phi=90^\circ$),
the thermal fluctuation in $\phi$ incidentally breaks the symmetry to provide finite values of excitonic coupling,
realizing the ultrafast exciton transfer between the two dps
via the thermal dynamics in $Q$ that frequently pass through $Q=0$.

It would be insightful to estimate the exciton transfer rate based on
the averaged $\left|V\right|^2$, i.e.,
$\Braket{\left|V\right|^2}=1.29\times10^4\,{\rm cm^{-2}}$ (see \ref{phi_and_V}), via FGR.
Assuming classical nuclei and taking the high-temperature limit,
the FGR expression of the exciton transfer rate reduces to the well-known Marcus formula
\begin{align}
k_{\rm FGR} = \frac{2\pi}{\hbar}\Braket{\left|V\right|^2}
\frac{1}{\sqrt{4 \pi \lambda k_{\rm B} T}}
\exp\left( -\frac{\left(\Delta G^\circ + \lambda\right)^2}{4 \lambda k_{\rm B} T} \right)
\label{marcus}
\end{align}
where the reaction Gibbs free energy change $\Delta G^\circ$ is zero by symmetry.
$\lambda$ denotes the reorganization energy, and its value can be estimated from Figure \ref{FIG6}b as $28.6\,{\rm meV}$.
$T$ is the temperature and set to $298.15\,{\rm K}$.
According to these parameters and eq \ref{marcus}, the FGR rate is calculated as $k_{\rm FGR}=1.5\times10^{13}\,{\rm s^{-1}}$,
 leading to the time constant of $\tau_{\rm FGR}=6.7\times10^{-14}\,{\rm s}$.
Remarkably, this is in the same orders of magnitude
with the time constant obtained from the NA-MD simulations ($\tau=3\times10^{-14}\,{\rm s}$).

It is worth to point out the analogy between the present exciton-transfer mechanism
and the proton-coupled electron transfer (PCET) mechanism.
The PCET process is described as a diabatic transition along the reaction coordinate, which is
a fast nuclear motion, i.e., the movement of H atom, where the diabatic coupling is dependent on
the slow nuclear degree of freedom, i.e., the solvent coordinate.\cite{Azzouz1993,HammesSchiffer2010,Sirjoosingh2011,Mandal2019}
$Q$ and $\phi$ in the present mechanism can be seen as analogues of the fast and slow nuclear degree of freedom
in the PCET mechanism, respectively.

\section{CONCLUSION}
To summarize, the inter-ligand exciton transfer dynamics in \ce{Zn(dp)2} complex was investigated focusing on
the effect of dynamic fluctuation in excitonic coupling $V$.
The NA-MD simulations were used in combination with exciton density analysis to characterize the exciton transfer,
and the results were interpreted in the context of two-state model.
The temporal change in the electronic-state population obtained from the NA-MD simulations
highlighted the ultrafast exciton transfer between the two dps,
being qualitatively consistent with the experimental observation.
The exciton transfer events in the NA-MD simulations are classified into ``adiabatic'' and ``non-adiabatic'' transfers,
which occurs without and with a non-adiabatic transition, respectively,
where the former was the dominant mechanism.
The excitonic coupling was found to have a strong correlation with the displacement of dihedral angle between
the two dps from 90$^\circ$.
In fact, the distribution of dihedral angle indicated that the exciton transfer tends to occur
when the relative orientation of two dp planes is not orthogonal.
The reaction coordinate for the exciton transfer was determined by finding the atomic displacement vector
that linearly coupled with the diabatic energy gap via a regression analysis.
The reaction coordinate involves primarily the stretching motion of the C--H bonds on the central carbon atoms
in dps.
There is also a partial contribution of bending motion and
C--C bond stretching between the central carbon atoms and the 5-membered rings.
The effect of the structural deformation along the reaction coordinate on the diabatic energy gap
can be qualitatively understood in terms of the relative phases in the HOMO and LUMO of the monomer dp.
The VACFs for the dihedral angle between the two dps and the reaction coordinate indicate that the
individual exciton transfer events, which is driven by the structural motion along the reaction coordinate,
is faster than the dihedral angle fluctuation that breaks the symmetry.
Both the dihedral angle and the reaction coordinate had Gaussian-shaped probability distribution, suggesting that
their dynamics are dominated by thermal fluctuation.
The reaction coordinate values at the moments of ``adiabatic'' exciton transfer events had considerably narrower
distribution than that of the entire trajectories.
This result shows that the exciton transfer tends to occur in the vicinity of the origin of reaction coordinate,
which, within the framework of two-state model, corresponds to the diabatic crossing point.
These findings suggest that the exciton transfer in the present system can be understood in the following mechanism;
the exciton transfer can occur at the moment that the system pass through the diabatic crossing point
via the atomic motion along the reaction coordinate,
but only when the excitonic coupling is ``turned on'' due to the incident twisting between the two dps.
Overall, the present work establishes the basic theoretical concept and computational approach to understand the
exciton transfer phenomena under the dynamic fluctuation in the excitonic coupling.

\section*{SUPPLEMENTARY MATERIAL}
The supplementary PDF file contains visualized NTOs, vertical and relaxed potential energies for each state, recrossing time distribution, detailed definition of $\phi$, detailed procedure for calculation of monomer excitation energies, normal-mode decomposition analyses of $\bar{\bf Q}$, regression analysis results using the S$_1$ stable geometry as origin, characterization of the ground-state dynamics.

\section*{ACKNOWLEDGMENT}
The authors thank Dr. Ryojun Toyoda and Prof. Ryota Sakamoto in Tohoku University, Japan, for fruitful discussion.
Financial supports were provided by Japan Science Technology Agency PRESTO (grant no. JPMJPR23Q2),
Toyota Physical and Chemical Research Institute, Japan, and Institute for Quantum Chemical Exploration, Japan.
A part of the computation was conducted using Research Center for Computational Science, Okazaki, Japan (Project: 25-IMS-C019).

\bibliography{reference}

\end{document}